\documentclass[aps,showpacs]{revtex4}

\usepackage{graphicx}

\def\spr{{s^{\prime}}}

\def\lmb{{\lambda_{\beta}}}
\def\pzp{p_0^{\prime}}
\def\pmp{p^{\prime}}
\def\pmpb{\mbox{\bfseries\itshape p}^{\prime}}
\def\pmb{\mbox{\bfseries\itshape p}}
\def\kmb{\mbox{\bfseries\itshape k}}
\def\eb{e_{\beta}}

\def\db{\partial_{\beta^2}}

\def\pp{{\bar p}}

\def\Spp{^3S_1^{+}}
\def\Dpp{^3D_1^{+}}

\def\Sp{^1S_0^{+}}

\def\im{{\rm Im\,}}
\def\re{{\rm Re\,}}

\begin{document}

\title{Separable Kernel of Nucleon-Nucleon Interaction
in the Bethe-Salpeter Approach}

\author{Serge G. Bondarenko}
\email{bondarenko@jinr.ru}
\author{Valery V. Burov}
\email{burov@thsun1.jinr.ru}
\affiliation{BLTP, Joint Institute for Nuclear Research,
141980 Dubna, Russia}

\author{Naohide Hamamoto}
\email{naohide@rcnp.osaka-u.ac.jp}
\author{Hiroshi Toki}
\email{toki@rcnp.osaka-u.ac.jp}
\affiliation{RCNP, Osaka University, Mihogaoka, Ibaraki, Osaka, 567-0047, Japan}

\begin{abstract}
The dispersion relations for nucleon-nucleon ($NN$) $T$-matrix in
the framework of the Bethe-Salpeter equation for a two spin-one-half
particle system and with a separable kernel of interaction are
considered. The developed expressions are applied for
construction of the separable kernel of interaction for $S$
partial-waves in singlet and triplet channels. We calculate the
low energy scattering parameters, the phase shifts, and the
deuteron binding energy with the separable interaction. The
approach can be easily extended to higher partial-waves for
$NN$-scattering and other reactions ($\bar N N$-, $\pi N$-
scattering).
\end{abstract}


\maketitle

\section{Introduction}

Recent and planned experiments on $NN$-scattering~\cite{lehar}
as well as reactions with the deuteron and light nuclei reach large
momentum transfer from the probe particle to the investigated
system. In this context, relativistic effects play an essential
role in understanding reaction mechanisms. Studies of elastic
electron-deuteron scattering (see, for example, \cite{GH,pair}) in
the nonrelativistic framework require to take into account
mesonic exchange currents (a part of them has a pure relativistic
origin, i.e. the so-called pair currents~\cite{BB97}). Theoretical
calculations should therefore follow relativistic approaches.

The approach based on the Bethe-Salpeter equation (BS
approach)~\cite{BS51} gives a powerful method to investigate
various nuclear reactions both for electromagnetic and strong
interactions. To find an analytic solution for the BS equation for
$T$-matrix of $NN$ scattering describing bound and
scattering states, the separable kernel of interaction is very
useful~\cite{yam}. The separable kernel is well known from
nonrelativistic consideration to solve the Lippman-Schwinger equation
analytically~\cite{pless}. The separable form of interaction is
good not only for the technical reason but, in our opinion,
also for the reflection of the nonlocal type of elementary nucleon-nucleon
interaction (it was stressed first in papers~\cite{yam}). This
allows us not only to avoid the divergence peculiar of local theories
but also to take into account the internal structure of the nucleon.

In the paper, we consider the BS equation for $T$-matrix and solve
it using the separable kernel of interaction for a spin-one-half
particle system (for a spin-zero case see, for example,
\cite{zingl} and for trinucleon systems~\cite{rupp-tjon}).
This allows us to solve the BS equation without
referring to the ladder approximation.   We also find dispersion
relations for the obtained solution which allow us to perform explicit
analytic calculations and connect internal parameters of the
kernel with experimental data: low-energy parameters, bound state
energy, and phase shifts.

The paper is organized as follows: after the partial-wave decomposition
of the BS equation (section~\ref{sec:decomp}), the one-rank separable kernel
of interaction is introduced and used for solving the BS equation
(section~\ref{sec:separable}). Dispersion relations for the $T$-matrix
are discussed in section~\ref{sec:disper}.
In section~\ref{sec:ladder}, it is shown that the separable kernel
of interaction includes the kernel in the ladder approximation, and the
connection between parameters of two kernels is found.
Section~\ref{sec:results} is devoted to results of calculation of
the low-energy parameters, bound state energy, and phase shifts
of elastic $NN$-scattering and to the discussion for the future works.

\section{Partial-wave decomposition of the BS equation}
\label{sec:decomp}

The BS equation for the two-nucleon $T$-matrix
with momenta $p_1 (p_1^{\prime}),
p_2 (p_2^{\prime})$ could be represented in the following form:
\begin{eqnarray}
T(p^{\prime}, p; P) = V(p^{\prime}, p; P) +
\frac{i}{(2\pi)^4}\int dk\, V(p^{\prime}, k; P) S(k; P) T(k, p; P),
\label{t00}
\end{eqnarray}
where $P=p_1+p_2$, $p=(p_1-p_2)/2$ [$p^{\prime}=(p_1^{\prime}-p_2^{\prime})/2$]
are total and relative momenta, and $V(p^{\prime}, p; P)$ is the $NN$ kernel
of interaction. The two-body propagator in terms of the total
and relative momenta has the form:
\begin{eqnarray}
S(k; P)=1/\left[({\hat P}/2+{\hat k}-m+i\epsilon)
({\hat P}/2-{\hat k}-m+i\epsilon)\right],
\end{eqnarray}
with ${\hat a} = a_{\mu} \cdot \gamma^{\mu}$, and $\gamma^{\mu}$ are
Dirac matrices.

After the partial wave decomposition in the rest frame of the
two-nucleon system, the equation~(\ref{t00}) becomes ($s=P^2$),
\begin{eqnarray}
T_{\alpha \beta}(\pzp, \pmp, p_0, p; s) &=&
V_{\alpha \beta}(\pzp, \pmp, p_0, p; s)
\label{t01}\\
&+& \frac{i}{4\pi^3} \sum_{\gamma \delta} \int\, dk_0\,\int\,
k^2\, dk\, V_{\alpha \gamma}(\pzp, \pmp, k_0, k; s) S_{\gamma
\delta}(k_0, k; s) T_{\alpha \beta}(k_0, k, p_0, p; s).
\nonumber\end{eqnarray}
(Here and below $p \equiv |\pmb|$, $\pmp
\equiv |\pmpb|$, $k \equiv |\kmb|$). Greek letters denote
partial waves in the channels under consideration. In
$LSJ\rho$-basis~\cite{KU72,bb:cov2}, the singlet $^1S_0$ channel
consists of four partial waves, and the triplet $^3S_1$ channel, of
eight partial-waves.

Below we omit all states which have at least one negative-energy
nucleon and take only $\Sp$-wave in the singlet channel
and neglect also $\Dpp$-wave and take only $\Spp$-wave in the triplet channel.
Therefore eq.~(\ref{t01}) could be rewritten as:
\begin{eqnarray}
t(\pzp, \pmp, p_0, p; s) = v(\pzp, \pmp, p_0, p; s)
\hskip 70mm
\label{t02}\\
+ \frac{i}{4\pi^3}
\int\, dk_0\,\int\, k^2\, dk\, \frac{v(\pzp, \pmp, k_0, k;s)\,
t(k_0, k, p_0, p; s)}{(\sqrt{s}/2-e_k+i\epsilon)^2-k_0^2},
\nonumber\end{eqnarray}
where $t \equiv T_{{\tilde \alpha}\, {\tilde \alpha}},
v \equiv V_{{\tilde \alpha}\, {\tilde \alpha}}$, with
${\tilde \alpha} = \Sp\mbox{ or }\Spp$, and $e_k = \sqrt{k^2+m^2}$
and $m$ is the nucleon mass.

The $T$-matrix normalization condition could be written in the on-mass-shell
form. Introducing phase shifts $\delta(s)$, one could write
\begin{eqnarray}
t(s) \equiv t(0,\pp,0,\pp,s) = - \frac{16 \pi}{\sqrt{s}\sqrt{s-4m^2}}\,
e^{i\delta(s)}\, \sin{\delta(s)},
\label{t03}\end{eqnarray}
with $\pp = \sqrt{s/4-m^2} = \sqrt{2mT_{lab}}$.

To introduce also the low-energy parameters of $NN$-scattering,
it is suitable to expand the above expression into a series of
$\pp$-terms with the following
expression~\cite{brown}:
\begin{eqnarray}
\pp\, {\cot}\, \delta(s) = - a_0^{-1} + \frac{r_0}{2}\pp^2 + {\cal O}(\pp^3).
\label{t03a}\end{eqnarray}
Taking into consideration only the first two terms
of the decomposition~(\ref{t03a}),
we define the scattering length  $a_0$ and the effective radius $r_0$.

As for the mass of the bound state ($M_b$), the squared $T$-matrix
has a simple pole in the total momentum squared ($s$),
and the bound state conditions could be written in the following form:
\begin{eqnarray}
t(\pzp, \pmp, p_0, p; s) = \frac{B(\pzp, \pmp, p_0, p; s=M_b^2)}{s-M_b^2}
+ R(\pzp, \pmp, p_0, p; s),
\label{t03b}\end{eqnarray}
where functions $B$ and $R$ are regular at the point $s=M_b^2$.
The bound state energy $E_b$ is connected with $M_b$: $M_b = 2m-E_b$.

\section{Rank I separable kernel of interaction}
\label{sec:separable}

The {\em ansatz} for the separable kernel of interaction with rank I has the form:
\begin{eqnarray}
v(\pzp, \pmp, p_0, p; s) = \lambda g(\pzp, \pmp) g(p_0, p).
\label{t04}\end{eqnarray} Here introduced is the g-function for
$NN$-vertices, and $\lambda$ is the coupling parameter.

To solve (\ref{t02}) we could assume the following separable form
for $T$-matrix:
\begin{eqnarray}
t(\pzp, \pmp, p_0, p; s) = \tau(s) g(\pzp, \pmp) g(p_0, p)
\label{t05}\end{eqnarray}
and substituting (\ref{t04}) and (\ref{t05}) into (\ref{t02}),
we obtain the function $\tau(s)$:
\begin{eqnarray}
\tau(s) = 1/(\lambda^{-1} + h(s)),
\label{t06}\end{eqnarray}
where the function $h(s)$ is:
\begin{eqnarray}
h(s) = -\frac{i}{4\pi^3}\, \int\, dk_0\,\int\, k^2\, dk\,
\frac{[g(k_0,k)]^2}{(\sqrt{s}/2-e_k+i\epsilon)^2-k_0^2}.
\label{t07}\end{eqnarray}

Therefore, the $T$-matrix could be rewritten in the following form:
\begin{eqnarray}
t(\pzp, \pmp, p_0, p; s) = \frac{g(\pzp, \pmp) g(p_0, p)}
{\lambda^{-1} + h(s)},
\label{t08}\end{eqnarray}
and the on-mass-shell expression is:
\begin{eqnarray}
t(s) = \frac{n(s)}{d(s)} = \frac{[g(0, \pp)]^2}{\lambda^{-1} +
h(s)}. \label{t09}\end{eqnarray}
It should be noted that the
expression~(\ref{t09}) has the so-called $N/D$-form widely used in the
nonrelativistic $T$-matrix theory~\cite{brown} and in some
relativistic methods.

Using (\ref{t09}), it is easy to connect the solution of
$T$-matrix and the phase shifts $\delta(s)$. To achieve this, we
assume that the imaginary part of the function $n(s)$ satisfies  the
following condition:
\begin{eqnarray}
\im{} n(s) = 0. \label{t010}\end{eqnarray}
This condition is
connected with the specific choice of g-functions for
$NN$-vertices to be discussed in the next section.
By using (\ref{t09}) and (\ref{t010}), the phase shifts
$\delta(s)$ could be expressed by the formula:
\begin{eqnarray}
{\cot}\, \delta(s) = \frac{\re t(s)}{\im t(s)} = -\,
\frac{\lambda^{-1} + \re h(s)}{\im h(s)}
\label{t011}\end{eqnarray}

To express the low-energy parameters in terms of the $T$-matrix solution,
it is suitable to expand the function $h(s)$ in a series of $\pp$ terms:
\begin{eqnarray}
&& h(s) = h_0 + i\pp h_1 + \pp^2 h_2 + i \pp^3 h_3 + {\cal O}(\pp^4),
\label{t012a}\\
&& \re h(s) = h_0 + \pp^2 h_2 + {\cal O}(\pp^4),
\label{t012b}\\
&& \im h(s) = \pp (h_1 + \pp^2 h_3 + {\cal O}(\pp^3)).
\label{t012}\end{eqnarray}
Using now the definition (\ref{t03a})
and (\ref{t011}-\ref{t012}), we obtain the $a_0$ and $r_0$
parameters:
\begin{eqnarray}
a_0 = \frac{h_1}{\lambda^{-1}+h_0},\qquad r_0 =
\frac{2}{h_1}\left[(\lambda^{-1}+h_0)\frac{h_3}{h_1}-h_2\right].
\label{t013}\end{eqnarray}
The bound state condition~(\ref{t03b})
with the help of (\ref{t08}) is now expressed in the form:
\begin{eqnarray}
\lambda^{-1} = -h(s=M_b^2).
\label{t014}\end{eqnarray}

\section{Dispersion relations}
\label{sec:disper}

Let us consider now the analytic properties of the BS solution
(namely, the function $h$).
The simplest choice of the function $g(p_0,p)$ is the {\em Yamaguchi}
type~\cite{yam}:
\begin{eqnarray}
g(p_0,p) = (p_0^2-p^2-\beta^2+i\epsilon)^{-1}.
\label{t21a}\end{eqnarray}

In this case, the function $h$ could be rewritten as
\begin{eqnarray}
h(s,\beta) = -\,\frac{i}{4\pi^3}\, \db\, \int dp_0\,\int\, p^2\, dp\
\frac{1}{(\sqrt{s}/2-e_p+i\epsilon)^2-p_0^2}\ \frac{1}{p_0^2-\eb^2+i\epsilon}
\label{t21}\end{eqnarray}
with $\db \equiv \partial / \partial \beta^2$ and $\eb=\sqrt{p^2+\beta^2}$.
We have introduced the second argument $\beta$ to extract the explicit
dependence of the function $h$ on this parameter.

Analyzing equation~(\ref{t21}), we could conclude that there are four poles in
the complex plane $p_0$, namely:
\begin{eqnarray}
&& p^{(1)}_0(s) = \frac{\sqrt{s}}{2} - e_p + i\epsilon \qquad\qquad
p^{(2)}_0(s) = - \frac{\sqrt{s}}{2} + e_p - i\epsilon
\label{t22}\\
&& p^{(3)}_0(s) = - \eb + i\epsilon \qquad\qquad\quad\quad
p^{(4)}_0(s) = \eb - i\epsilon
\nonumber\end{eqnarray}

With the change of the variable $s$, the poles $p^{(1)}_0$ and
$p^{(3)}_0$ move, and there could be a situation when two poles
``pinch'' real $p_0$ axes. It means, the function $h(s)$ has at
this $s$-point the leap and imaginary part. First points at which
this condition is satisfied (branch points) could be found from
the following equations:
\begin{eqnarray}
&& p^{(1)}_0(s) = - p^{(1)}_0(s) \qquad \Rightarrow \qquad s_0 = 4m^2,
\label{t23}\\
&& p^{(1)}_0(s) = - p^{(2)}_0(s) \qquad \Rightarrow \qquad s_1 = 4(m+\beta)^2.
\nonumber\end{eqnarray}

Summarizing the situation, one could say that the function $h(s)$ has two
cuts starting at points $s_0$ and $s_1$, respectively,
and therefore could be written in a dispersion form:
\begin{eqnarray}
&& h(s,\beta) = \int\limits_{4m^2}^{+\infty}
\frac{\rho(s^{\prime},\beta)\,ds^{\prime}}{s^{\prime}-s-i\epsilon},
\label{t24}\\
&& \rho(s^{\prime},\beta) = \theta (s^{\prime}-4m^2)
\rho_{el}(s^{\prime},\beta) + \theta
(s^{\prime}-4(m+\beta)^2)\rho_{in}(s^{\prime},\beta)
\nonumber\end{eqnarray} with two spectral functions $\rho_{el;in}$
({\em el} stands for {\em elastic} and {\em in} for {\em
inelastic}) which are connected with the imaginary parts as
\begin{eqnarray}
\rho(s^{\prime},\beta) = \frac{1}{\pi} \im h(s^{\prime},\beta) =
\frac{1}{2\pi i} (h-h^{*}).
\label{t25}\end{eqnarray}

To find spectral functions, one should  perform $p_0$-integration
in (\ref{t21}) which gives the following result:
\begin{eqnarray}
h(s,\beta) = -\frac{1}{2\pi^2}\,\db\, \int p^2\, dp\,
\frac{1}{s/4-\sqrt{s}e_p+m^2-\beta^2+i\epsilon}
\left[\frac{1}{\sqrt{s}-2e_p+i\epsilon}+\frac{1}{2\eb}\right].
\label{t26}\end{eqnarray}
Taking into account the following symbolic equation:
\begin{eqnarray}
\frac{1}{x-x_0 \pm i\epsilon} = \frac{{\cal P}}{x-x_0} \mp i\pi\delta(x-x_0),
\label{t27}\end{eqnarray}
it is easy to find the spectral functions:
\begin{eqnarray}
&& s^{\prime} \ge 4m^2
\label{d28a}\\
&& \hskip 10mm
\rho_{el}(s^{\prime},\beta) =
\sqrt{\spr}\sqrt{\spr-4m^2}/(\pi^2(\spr-4m^2+4\beta^2)^2)
\\
&& \spr \ge 4(m+\beta)^2
\label{d28b}\\
&& \hskip 10mm
\rho_{in}(\spr,\beta) =
-(64\beta^6+16\beta^4\spr-192\beta^4m^2-20\beta^2\spr^2
\nonumber\\
&& \hskip 10mm
+192\beta^2m^4-32\beta^2\spr m^2+16m^4\spr+3\spr^3-64m^6-12m^2\spr^2)/
\nonumber\\
&& \hskip 10mm
(2\pi^2\spr(\spr-4m^2+4\beta^2)^2\sqrt{\spr-4(m+\beta)^2}
\sqrt{\spr-4(m-\beta)^2}).
\nonumber\end{eqnarray}

To perform integration~(\ref{t21}), it is suitable to introduce new variables:
\begin{eqnarray}
&& \lmb = \frac{\beta}{m},\quad\quad\qquad\qquad\qquad t=\frac{s}{4m^2}
\label{t211a}\\
&& \rho^2 = 1 - \frac{4m^2}{s} = 1 - \frac{1}{t},\qquad
w^2 = \frac{m^2}{m^2-\beta^2} = \frac{\lmb^2}{1-\lmb^2},
\label{t211b}\\
&& v^2=\frac{m+\beta}{m-\beta}= \frac{1+\lmb}{1-\lmb},
\qquad
\sigma^2 = - \frac{s - 4(m+\beta)^2}{s - 4(m-\beta)^2}=
-\frac{t-(1+\lmb)^2}{t-(1-\lmb)^2}.
\label{t211c}
\end{eqnarray}

In case when the following conditions are valid
\begin{eqnarray}
&& m > \beta > 0 \qquad \Rightarrow \qquad 1> \lmb > 0, \\
\label{t212a}
&& 4(m+\beta)^2 > s >4m^2 \qquad \Rightarrow \qquad (1+\lmb)^2 > t > 1,
\label{t212b}\end{eqnarray}
the introduced parameters are real and positive:
\begin{eqnarray}
1 > \rho^2 > 0,
\qquad\qquad
w^2 > 0,
\\
v^2 > 1 > 0,
\qquad\qquad
\sigma^2 > 0.
\nonumber\end{eqnarray}
The condition~(\ref{t212b}) means that the second (inelastic) imaginary part
does not contribute to the function $\im h(s)$
when the phase shifts are calculated in the region $4(m+\beta)^2 > s >4m^2$,
and therefore:
\begin{eqnarray}
\im h(s,\beta) = \im h_{el}(s,\beta) =
\frac{\rho(\rho^2-1)(1+w^2)^2}{4m^2\pi(\rho^2+w^2)^2},\quad\quad
\mbox{ if }\quad 4(m+\beta)^2 > s >4m^2.
\label{t213}\end{eqnarray}
Performing integration~(\ref{t24}), we
obtain for the real part of function $h(s)$ (the imaginary part is
given by (\ref{t213})):
\begin{eqnarray}
\re h(s,\beta) &=& h_{el}(s,\beta) + h_{in}(s,\beta),
\label{t215a}
\\
h_{el}(s,\beta) &=&
\label{t215b}
\frac{\rho(\rho^2-1)(1+w^2)^2}
{4m^2\pi^2(\rho^2+w^2)^2}
\ln{\left|\frac{\rho+1}{\rho-1}\right|}
\nonumber\\
&&
-\frac{(\rho^2-1)(1+w^2)^2(w^2-\rho^2)}{4m^2\pi^2(\rho^2+w^2)^2w}\arctan{1/w}
\nonumber\\
&&
-\frac{(\rho^2-1)(1+w^2)}{4m^2\pi^2(\rho^2+w^2)},
\nonumber\\
h_{in}(s) &=&
\label{t215c}
-\frac{(1+\sigma^2)(1+v^2)^2}{32m^2\pi^2(v^4+\sigma^2)}
\ln{\left|\frac{v^2+1}{v^2-1}\right|}
\nonumber\\
&&
-\frac{(1+\sigma^2)(1+v^2)^2}{16m^2\pi^2(v^2-1)^2v(v^2-\sigma^2)^2}
\nonumber\\
&&
\times
(v^6+v^4\sigma^2-4v^4+4v^2\sigma^2-v^2-\sigma^2)\arctan{1/v}
\nonumber\\
&&
+
\frac{(1+\sigma^2)(1+v^2)^3}{16m^2\pi^2\sigma(v^4+\sigma^2)(v^2-\sigma^2)^2\pi^2(v^2-1)^2}
\nonumber\\
&&
\times
(-v^6+2v^6\sigma^2-3v^4\sigma^2+3\sigma^4v^2-2\sigma^4+\sigma^6)\arctan{1/\sigma}
\nonumber\\
&&
-\frac{(1+\sigma^2)(1+v^2)^2}{16m^2\pi^2(v^2-1)(v^2-\sigma^2)}
\nonumber\end{eqnarray}

To find also expressions for low-energy parameters, we should return to
(\ref{t012})-(\ref{t013}) and expand the function $h(s)$ in a series of $\pp$ terms:
\begin{eqnarray}
h_0(\beta) &=&
\frac{(1+w^2)}{4m^2\pi^2w^3}(w+(1+w^2)\arctan{1/w})
\label{t216a}\\
&&
+
\frac{5+20v^2+5v^8+14v^4+20v^6}{4m^2\pi^2(v^2-1)^3(3v^2+1)}
\sqrt{\frac{3v^2+1}{v^2+3}}\arctan{1/\sqrt{\frac{3v^2+1}{v^2+3}}}
\nonumber\\
&&
-\frac{(v^2+1)^4}{4m^2\pi^3v(v^2-1)^3}\arctan{1/v}
-\frac{1}{8m^2\pi^2}\ln{\left|\frac{v^2+1}{v^2-1}\right|}
-\frac{(v^2+1)^2}{4m^2\pi^2(v^2-1)^2}
\nonumber\\
h_2(\beta) &=&
-
\frac{(1+w^2)^2}{4m^4\pi^2w^5}(3w+(3+w^2)\arctan{1/w})
\label{t216b}\\
&&
-\frac{1}{4m^4\pi^2(v^2-1)^5(v^2+3)(3v^2+1)^2}
(304v^2+1212v^{12}+3790v^8
\nonumber\\
&&
+304v^{14}+2704v^{10}+29v^{16}+1212v^4+
2704v^6+29)
\nonumber\\
&&
\times
\sqrt{\frac{3v^2+1}{v^2+3}}
\arctan{1/\sqrt{\frac{3v^2+1}{v^2+3}}}
\nonumber\\
&&
+
\frac{(v^4+10v^2+1)(v^2+1)^4}{4m^4\pi^2v(v^2-1)^5}\arctan{1/v}
\nonumber\\
&&
+
\frac{1}{8m^4\pi^2}\ln{\left|\frac{v^2+1}{v^2-1}\right|}
+\frac{11v^8+52v^6+66v^4+52v^2+11)(v^2+1)^2}{8m^4\pi^2(v^2-1)^4(3v^2+1)(v^2+3)}
\nonumber\\
h_1(\beta) &=& \frac{(1+w^2)^2}{4m^3\pi w^4}
\label{t216c}\\
h_3(\beta) &=& -\frac{(1+w^2)^2(4+3w^2)}{8m^5\pi w^6}
\label{t216d}\end{eqnarray}

At this point we have an{\em explicit analytic} expressions which connect
the parameters of the separable kernel of interaction ($\lambda$ and
$\beta$ [$\lmb$]) with observables:
phase shifts $\delta(s)$ (eq.~(\ref{t011})), low-energy parameters
$a_0$ and $r_0$ (eq.~(\ref{t013})), and bound state energy
(eq.~(\ref{t014})).

\section{Separable and one-meson exchange kernels of interaction}
\label{sec:ladder}

In this section, we show that the so-called realistic kernel of interaction
(meson-nucleon) in the ladder approximation is included in the separable kernel
of the Yamaguchi-type g-functions.
To achieve this, we introduce a simple approximation for the ladder kernel:
\begin{eqnarray}
V(p_0,p;k_0,k) \to {\tilde V}(p_0,p;k_0,k) =
\frac{V(p_0,p;0,0)V(0,0;k_0,k)}{V(0,0;0,0)}.
\label{ns01}
\end{eqnarray}
Using the expressions of the kernel for scalar-meson (sc)
exchange~\cite{KU72},\cite{bonn}:
\begin{eqnarray}
V_{sc}(p_0,p;k_0,k) =
-\frac{g_{sc}^2}{4\pi}\frac{1}{\pi^2}\frac{1}{4pke_pe_k}
\left[(e_pe_k+m^2)Q_0(z_{\mu})-pkQ_1(z_{\mu})\right],
\label{ns02}
\end{eqnarray}
where $z_{\mu}=(p^2+k^2-(p_0-k_0)^2+\mu^2)/(2pk)$, $\mu$ is
the exchange meson mass, and $Q_{i}(z)$ are the Legendre functions of second
kind, $Q_0(z) = 1/2\ln{(z+1)/(z-1)}$, $Q_1(z) = zQ_0(z) - 1$, it
is clear that
\begin{eqnarray}
V_{sc}(p_0,p;0,0) = \lim_{k_0\to 0, k\to 0} V_{sc}(p_0,p;k_0,k)
= a_p g_{\mu}(p_0,p),
\label{ns03}
\end{eqnarray}
with
\begin{eqnarray}
a_p = \frac{g_{sc}^2}{4\pi}\frac{1}{\pi^2}\frac{e_p+m}{2e_p},
\nonumber\\
g_{\mu}(p_0,p) = 1/(p_0^2-p^2-\mu^2).
\label{ns05}
\end{eqnarray}
Expressions for $V_{sc}(0,0;k_0,k)$ could be obtained
from eqs.~(\ref{ns03}-\ref{ns05}) by the following substitutions:
$p_0 \to k_0$ and $p \to k$. To make connection between parameters more clear,
we perform the $p/m$-decomposition in function $a_p$ up to ${\cal O}(p^2/m^2)$
term:
\begin{eqnarray}
a_p = a_k = \frac{g_{sc}^2}{4\pi}\frac{1}{\pi^2},
\nonumber\\
V_{sc}(0,0;0,0) = -\frac{g_{sc}^2}{4\pi}\frac{1}{\pi^2}\frac{1}{\mu^2}.
\label{ns06}
\end{eqnarray}

Using now expression~(\ref{ns01}), we could write:
\begin{eqnarray}
{\tilde V}_{sc}(p_0,p;k_0,k) = -\frac{g_{sc}^2}{4\pi}(\frac{\mu}{\pi})^2
g_{\mu}(p_0,p) g_{\mu}(k_0,k).
\label{ns07}
\end{eqnarray}

Comparing the above expressions with the separable form of the kernel
introduced by (\ref{t04}),
we could write the following connection between parameters:
\begin{eqnarray}
\beta = \mu,\qquad\qquad
\lambda_{sc} = -\frac{g_{sc}^2}{4\pi}(\frac{\mu}{\pi})^2.
\label{ns08}
\end{eqnarray}

Equations~(\ref{ns08}) are valid also for the vector-meson (vc) exchange
kernel with the substitution of $g_{sc} \to g_{vc}$
and $\mu$ by the vector-meson mass
and could be derived from the following expression:
\begin{eqnarray}
V_{vc}(p_0,p;k_0,k) =
-\frac{g_{vc}^2}{4\pi}\frac{1}{\pi^2}\frac{1}{4pke_pe_k}
\left[-2(2e_pe_k+m^2)Q_0(z_{\mu})\right].
\label{ns09}
\end{eqnarray}

The pseudoscalar-meson (ps) exchange kernel of interaction has the
form:
\begin{eqnarray}
V_{ps}(p_0,p;k_0,k) =
-\frac{g_{sc}^2}{4\pi}\frac{1}{\pi^2}\frac{1}{4pke_pe_k}
\left[-(e_pe_k-m^2)Q_0(z_{\mu})+pkQ_1(z_{\mu})\right],
\label{ns10}
\end{eqnarray}
and for $p,k \to 0$ we could write:
\begin{eqnarray}
{\tilde V}_{ps}(p_0,p;k_0,k) \sim p^2 k^2 g_{\mu}(p_0,p) g_{\mu}(k_0,k).
\label{ns11}
\end{eqnarray}
In this case, the expression for ${\tilde V}_{ps}$ goes to zero for
$p,k \to 0$, and it is impossible to find a connection between parameters
similar to (\ref{ns08}).

We have obtained connections between parameters for various cases of
the scalar and vector exchange, which are given in Ref.~\cite{bonn}.
Results are given in Table~\ref{ns12}.

\begin{table}[htbp]
\caption{\label{ns12} Connection between parameters of two kernels.}
\begin{center}
\begin{tabular}{lllll}
\hline\hline
$J^P$ & & $\mu$ (GeV) & $g^2/4\pi$ & $\lambda$ (GeV$^{2}$) \\
\hline
$0^+$ & $NN\delta$ & 0.983 & 0.64 & -.06265954891 \\
$0^+$ & $NN\sigma^{\prime}$ & 0.550 & 7.07 & -.2166930823 \\
$1^-$ & $NN\rho$ & 0.769 & 0.43 & -.02576448048 \\
$1^-$ & $NN\omega$ & 0.7826 & 10.6 & -.6577877886\\
\hline\hline
\end{tabular}
\end{center}
\end{table}

\section{Results and discussion}
\label{sec:results}

At this moment, we could find internal parameters of the kernel of
interaction ($\lambda$, $\beta$) to reproduce experimental values
for low-energy parameters $a_{0s}^{exp}, r_{0s}^{exp}$ for the singlet
channel ($^1S_0$) and $a_{0t}^{exp}$ and bound state (deuteron)
energy $E_d^{exp}$ for the triplet channel ($^3S_1$).

{\bfseries In the case of $^1S_0$-channel}, we use
equation~(\ref{t013}) with $a_{0} \equiv a_{0s}^{exp}$ to find
$\lambda$:
\begin{eqnarray}
{\lambda}^{-1} = (a_{0s}^{exp})^{-1} h_1(\beta) - h_0(\beta).
\label{dis01}
\end{eqnarray}
Inserting the obtained expression into equation~(\ref{t014}) with $r_0
\equiv r_{0s}^{exp}$, we find:
\begin{eqnarray}
r_{0s}^{exp} = \frac{2}{h_1(\beta)} \left[ (a_{0s}^{exp})^{-1}
h_3(\beta) - h_2(\beta)\right]. \label{dis02}
\end{eqnarray}
Solving the nonlinear equation (\ref{dis02}), we could find the value of
$\beta$ and, then, using expression~(\ref{dis01}), the value of
$\lambda$.

{\bfseries In the case of $^3S_1$-channel}, we obtain $\lambda$
from the bound state condition~(\ref{t014}) with $E_b \equiv
E_d^{exp}$:
\begin{eqnarray}
\lambda^{-1} = - h(s=(M_d^{exp})^2,\beta), \label{dis03}
\end{eqnarray}
where $M_d = 2m-E_d$. Inserting the obtained expression into
(\ref{t013}) with $a_0 \equiv a_{0t}^{exp}$, we find:
\begin{eqnarray}
a_{0t}^{exp} = \frac{h_1(\beta)}{h_0(\beta) - h(s=M_d^2,\beta)}.
\label{dis04}
\end{eqnarray}
Solving the nonlinear eq.~(\ref{dis04}), we could find the value of
$\beta$ and, then, using expression~(\ref{dis03}) - the value of
$\lambda$.

To solve equations~(\ref{dis02}) and (\ref{dis04}), we use FORTRAN
code with the DZEROX subroutine (as a part of CERNLIB package).
As a result, we find the following values for $\lambda$ and $\beta$:
\begin{eqnarray}
\begin{array}{ll}
\mbox{for $^1S_0$ channel}: & \\
& \lambda = -0.29425404 \mbox{ GeV}^{-2},
\qquad \beta = 0.22412880 \mbox{ GeV},\\
&\\
\mbox{for $^3S_1$ channel}: & \\
& \lambda = -0.79271213 \mbox{ GeV}^{-2},
\qquad \beta = 0.27160579 \mbox{ GeV},\\
\end{array}
\end{eqnarray}

The results on the binding energy of the deuteron and the low-energy
parameters are shown in Table~\ref{table3}. The experimental
data are taken from~\cite{Kroll}.

\begin{table}[htbp]
\caption{\label{table3} The binding energy and low-energy parameters
for singlet and triplet channels.}
\begin{center}
\begin{tabular}{llll}
\hline\hline
$1S_0$ & $a_{0s}$ (Fm) & $r_{0s}$ (Fm) & \\
\hline
Calculated & -23.748 & 2.75 & \\
Experiment & -23.748 $\pm$ 0.010 & 2.75 $\pm$ 0.05 & \\
\hline\hline
$3S_1$ & $a_{0t}$ (Fm) & $r_{0t}$ (Fm) & $E_d$ (MeV) \\
\hline
Calculated & 5.424 & 1.774 & 2.224644 \\
Experiment & 5.424 $\pm$ 0.004 & 1.759 $\pm$ 0.005 & 2.224644 $\pm$ 0.000046 \\
\hline\hline
\end{tabular}
\end{center}
\end{table}

\begin{figure}[ht]
\centerline{
\includegraphics[width=110mm,angle=270]{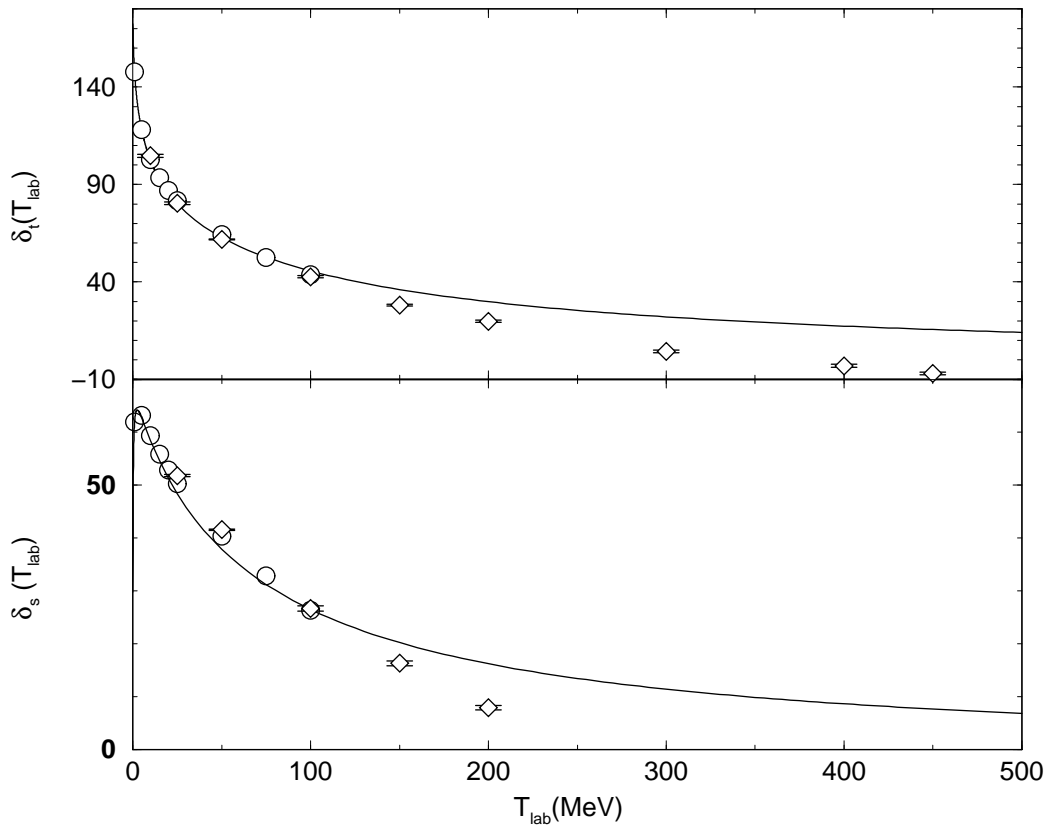}}
\caption{\label{fig1} The singlet $\delta_s$ and triplet
$\delta_t$ phase shifts.}
\end{figure}

The phase shifts $\delta_s(s)$ and $\delta_t(s)$ calculated with
the above parameters are presented in Fig.~\ref{fig1}. The experimental data
are taken from \cite{arndt}. As it is seen from
the figure, even with the simplest choice of the separable kernel of
interaction - rank I with only two parameters $\lambda$ and
$\beta$, we find the low-energy parameters of elastic $NN$
scattering $a_s$ and $r_s$ in the singlet channel and $a_t$ and
bound state (deuteron) energy $E_d$ in the triplet channel with required
accuracy and reproduce phase shifts in a region till $T_{lab} =
100-150$ MeV.

To summarize, we recall once again the basic points of our lines of
thought and the approximations made in the analysis. We start with
the Bethe-Salpeter equation for $T$-matrix and perform the
partial-wave decomposition. At this point, we omit all states which
have at least one negative-energy nucleon and consider only
$\Sp$-wave in the singlet channel and neglect also $\Dpp$-wave and
left only $\Spp$-wave in the triplet channel. In order to find
the explicit analytic solution for $T$-matrix, we introduce
separable ansatz for the kernel of interaction. As a result, we
find the $N/D$-form for on-shell $T$-matrix. Taking the simplest
choice of g-function in a {\em Yamaguchi} form for $NN$-vertex, we
analyze the analytic structure of the obtained solution for
$T$-matrix and write the dispersion representation of the
$T$-matrix. Performing the integration in the dispersion formula, we
obtain {\em explicit analytic} expressions which connect the
parameters of the separable kernel of interaction ($\lambda$ and
$\beta$ [$\lmb$]) with observables: phase shifts, low-energy
parameters and bound state energy. This connection is used to
calculate intrinsic parameters of the kernel of interaction and
then to find phase shifts.

We find also that the kernel of interaction in the ladder approximation is
included into the separable kernel. To obtain this result, we use a
simple approximation for the expressions in the one-meson exchange model.

The main result of the paper is the dispersion form of $T$-matrix
for elastic $NN$ scattering obtained for the separable kernel of
interaction which allows us to perform analytic calculations and to
connect explicitly parameters of the kernel and observables. The
discussed method could be technically expanded to the multi-rank
separable kernels and more complex forms of g-function for
$NN$-vertex to achieve required accuracy for bound state energy,
low-energy parameters and phase shifts in a wide region of cm laboratory
energy. The approach can be easily extended to higher partial-waves
for $NN$-scattering and other reactions ($\bar N N$-, $\pi N$- scattering).

{\bfseries Acknowledgments.}
We would like to thank Prof. Y. Yamaguchi and Prof. Y. Nambu
for useful discussions. One of us (V.V.B.)
expresses his deep gratitude to the directorate of RCNP for
hospitality.
The work was supported in part by the RFBR grant {\em N} 00-15-96737.


\begin{thebibliography}{100}

\bibitem{lehar} C. Lechanoine-LeLuc, F. Lehar,
Rev. of Modern Phys., {\bfseries 65} (1993), p. 47.

\bibitem{GH}      Gari M., Hyuga H.,
                  Z.Phys. A {\bfseries 58} (1976), p.291.

\bibitem{pair} V.V. Burov, V.N. Dostovalov, and S.Eh.Sus'kov, Czech. J. of
  Phys. {\bf 41} (1991) p. 1139; Particles and Nuclei
           {\bf 23} (1992), p.721.

\bibitem{BB97}
S.G.~Bondarenko, V.V.~Burov, M.~Beyer, S.M.~Dorkin,
Phys. Rev. C {\bfseries 58} (1998) 3143-3152.

\bibitem{BS51} E.E.~Salpeter and H.A.~Bethe,
Phys. Rev. C {\bfseries 84} (1951) p.1232.

\bibitem{yam} Y. Yamaguchi, Phys. Rev. {\bfseries 95}, (1954) p.1628;
Y. Yamaguchi, Y. Yamaguchi, Phys. Rev. {\bfseries 95}, (1954) p.1635.

\bibitem{pless}
L.~Mathelitsch, W.~Plessas, W.~Schweiger,
Phys. Rev. C {\bfseries 26} (1982) p.65;
W.~Schweiger, J.~Haidenbauer, W.~Plessas,
Phys. Rev. C {\bfseries 32} (1985), p.1261.

\bibitem{rupp-tjon}
G. Rupp and J.A. Tjon, Phys. Rev. C37, 1729 (1988);
G. Rupp and J.A. Tjon, Phys. Rev. C41, 472 (1990);
G. Rupp, Nucl. Phys. A508, 131c (1990);
G. Rupp and J.A. Tjon, Phys. Rev. C45, 2133 (1992);
G. Rupp, Proc. of PAN XIII, 13th Int. Conf. on Particles and
Nuclei, Perugia, June-July 1993,
World Scientific, Ed. A. Pascolini, p. 673-675 (1994).

\bibitem{zingl} K. Schwarz {\em et al.}, Acta Physica Austriaca,
{\bfseries 53} (1981), p.191;
J. Fr\" ohlich {\em et al.}, Phys. Rev. C
{\bfseries 25} (1982), p.2591;
 J. Fr\" ohlich {\em et al.}, Phys. Rev. C
{\bfseries 27} (1983), p.265;
J. Haidenbauer, and J. Froehlich, Phys. Rev. C{\bfseries 33}, 456 (1986).

\bibitem{KU72} J.J.~Kubis,
Phys. Rev. D{\bfseries 6} (1972) p.547.

\bibitem{bb:cov2}
S.G.~Bondarenko, V.V.~Burov, M.~Beyer, S.M.~Dorkin,
Few Body Syst. {\bfseries 26} (2-4) (1999) p.185.

\bibitem{brown} G.E. Brown, A.D. Jackson, ``The Nucleon-nucleon interaction'',
North-Holland Publishing Conpany, 1976.

\bibitem{bonn}  R. Machleidt {\em et al.},
Phys. Rep. {\bfseries 149} (1987) p.1.

\bibitem{Kroll} O. Dumbrajs {\em et al.},
Nucl. Phys. B {\bfseries 216} (1983) p.277.

\bibitem{arndt}
R. A. Arndt, L. D. Roper, R. A. Bryan, R. B. Clark,
B. J. VerWest, and P. Signell, Phys. Rev. D {\bfseries 28} (1983)  p.97;
 R.A.~Arndt, J.S.~Hyslop and L.D.~Roper,
Phys.Rev. D {\bfseries 35} (1987) p.128.


\end{thebibliography}
\end{document}